\documentclass{ws-p8-50x6-00}

\begin{document}

\title{A grid of Chemical Evolution Models along the Hubble Sequence}
\author{M. Moll\'{a} and A. I. D\'{\i}az}

\address{Universidad Aut\'{o}noma de Madrid,
       28049 Cantoblanco, Spain\\
E-mail: mercedes@pollux.ft.uam, angeles@pollux.ft.uam.es}

\author{and  F. Ferrini}

\address{Universit\`{a} di Pisa, Piazza Torricelli 2, 56100 Pisa, Italy\\
E-mail: federico@astro2.difi.unipi.it}  


\maketitle

\abstracts{
We have computed a grid of multiphase chemical evolution models whose
results are valid for any spiral galaxy, using as input the maximum
rotation velocity and the morphological type or index T}

\section{Introduction}

In previous works, we have calculated the chemical evolution of the
Solar Neighborhood (Ferrini et al~\cite{ferrini92}), the Galactic Disc
and Bulge (Ferrini et al~\cite{FMPD},\cite{mf95}), and a sample of
spiral galaxies (Moll\'{a} et al~\cite{MFD96,MHB99}) with the
multiphase model.  All of them reproduce with success the
observational data, in particular those related with the disk radial
distributions. They also predict the correlations
found along the Hubble Sequence (Moll\'{a} \& Roy~\cite{mr98}),
although the number of computed models is small to ensure it.  Now, we
have computed an extended grid of multiphase models, which allow the
comparison of observations with theoretical models for a large range
of galaxy masses and morphological types.

\section{The Multiphase Model Description} 

We assume a spherical protogalaxy with a gas mass which collapses to
fall onto the equatorial plane by forming the disk as a secondary
structure. The sphere is divided up into concentric cylindrical regions 1
kpc wide, with a {\sl halo} and a {\sl disk} region.  The Universal
Rotation Curve from Persic, Salucci \& Steel~\cite{PSS96}, an
analytical expression for V(R), is used to calculate the radial
distributions of the mass included in each of our cylinders $\rm
\Delta M$ (R).

The characteristic gas infall rate from the halo to the disk is
proportional to a parameter $f$, which, in turn, is inversely
proportional to the collapse timescale $\tau_{0}$, dependent on the
total mass (Gallagher et al~\cite{ght84}). It is easily computed
from its value for the MWG model, $\tau{\odot}$, and the ratio between
total masses: $\tau_{0}=\tau_{\odot}(M_{9},gal/M_{9,MWG})^{-1/2}$.

Stars form out in the halo, by a Schmidt law with a proportionality
factor $K$. In the disk, stars form in two steps:
molecular clouds form from the diffuse gas with a proportionality
factor $\mu$; then cloud-cloud collisions produce stars by a {\sl
spontaneous} process, at a rate proportional to a parameter
$H$. Moreover a {\sl stimulated} star formation process, 
by the interaction of massive stars with surrounding molecular cloud,
is assumed,s  proportional to a parameter $a$, 

These parameters are variable along the radius, by the volume effect,
through the use of the process efficiencies ($\epsilon$'s), valid for
the whole galaxy. We assume that the two efficiencies $\epsilon_{K}$
and $\epsilon_{a}$ are constant for all halos and galaxies, while
$\epsilon_{\mu}$, and $\epsilon_{H}$ change according the Hubble type
T by a function $\epsilon=\exp{(-T^{2}/a)}$. We have computed 10
types of galaxies for each mass.

\section{Preliminary Results and Conclusions}

A biparametric grid of models has been obtained for types 1 to 10 and
different luminosities. We obtain radial distributions for the star
formation rate, the diffuse and molecular gas densities and the
elemental abundances.

The most important result concerns the oxygen abundances: the radial
gradient only appears for the intermediate types ($7 < \rm{T} < 4$) at
all galaxy masses, being larger for the less massive
galaxies. However, the latest ones (T $ \ge 8$) have not developed a
gradient in a Hubble time, their abundances being $12 + \log{(O/H)}
\sim$ 7.5-8.  The early types T$ \leq 5$ reach a saturation level,
flattening the gradient once again, after having a steep one
early times.

This important result reproduces the observations and it solves the
apparent inconsistency of the largest gradients appearing in late type
spirals, while some irregulars shows no gradient at all.  The
explanation resides in the stimulated star formation which, being a
local process, maintains a minimum level of star formation constant
for all regions.


\begin{thebibliography}{99}

\bibitem{ferrini92}
Ferrini, F., Matteucci, F., Pardi, C., \& Penco, U., \Journal{ApJ}{387}{138} 
{1992}.

\bibitem{FMPD}
Ferrini, F., Moll\'{a}, M., Pardi, C., \& D\'{\i}az, A. I, \Journal{ApJ}{427} 
{745}{1994}.

\bibitem{ght84}
Gallagher, J. S., Hunter, D. A., \& Tutukov, A. V., \Journal{ApJ}{284}{544}
{1984}..

\bibitem{mf95}
Moll\'{a}, M. \& Ferrini, F. \Journal{ApJ}{454}{726}{1995}

\bibitem{MFD96}
Moll\'{a}, M., Ferrini, F., \& D\'{\i}az, A. I,\Journal{ApJ}{466}{668}{1996}. 

\bibitem{MHB99}
Moll\'{a}, M., Hardy, E., \& Beauchamp, D., \Journal{ApJ}{513}{695}{1999}.

\bibitem{mr98}
Moll\'{a}, M. \& Roy, J.-R., in {\em Chemical Evolution from zero to high
redshift}, eds. J.R.Walsh \& M.R. Rosa (Springer-Verlag, Berlin, 1999)

\bibitem{PSS96} 
Persic, M., Salucci, P. \& Steel, F., \Journal{MNRAS}{281}{27}{1996}.

\end{thebibliography}
\end{document}